\documentclass[pre,twocolumn,amsmath,amssymb,nofootinbib,floatfix,superscriptaddress]{revtex4}

\usepackage{graphicx,bm}
\usepackage[normalem]{ulem} 

\makeatletter
\def\graphicscale{\twocolumn@sw{0.3}{0.4}}
\def\graphicthreescale{\twocolumn@sw{0.3}{0.4}}

\begin{document}

\title{Three-dimensional ${\mathbb Z}_2$-gauge $N$-vector models}

\author{Claudio Bonati} \affiliation{Dipartimento di Fisica
  dell'Universit\`a di Pisa and INFN Sezione di Pisa, Largo Pontecorvo 3, I-56127 Pisa,
  Italy}

\author{Andrea Pelissetto}
\affiliation{Dipartimento di Fisica dell'Universit\`a di Roma Sapienza
        and INFN Sezione di Roma I, I-00185 Roma, Italy}

\author{Ettore Vicari} 
\affiliation{Dipartimento di Fisica dell'Universit\`a di Pisa,
        Largo Pontecorvo 3, I-56127 Pisa, Italy}

\date{\today}

\begin{abstract}
We study the phase diagram and critical behaviors of three-dimensional
lattice ${\mathbb Z}_2$-gauge $N$-vector models, in which an
$N$-component real field is minimally coupled with a ${\mathbb
  Z}_2$-gauge link variables. These models are invariant under global
O($N$) and local ${\mathbb Z}_2$ transformations. They present three
phases characterized by the spontaneous breaking of the global O($N$)
symmetry and by the different topological properties of the ${\mathbb
  Z}_2$-gauge correlations.  We address the nature of the three
transition lines separating the three phases.  The theoretical
predictions are supported by numerical finite-size scaling analyses of
Monte Carlo data for the $N=2$ model. In this case, continuous
transitions can be observed along both transition lines where the
spins order, in the regime of small and large inverse gauge coupling
$K$.  Even though these continuous transitions belong to the same $XY$
universality class, their critical modes turn out to be
different. When the gauge variables are disordered (small $K$), the
relevant order-parameter field is a gauge-invariant bilinear
combination of the vector field. On the other hand, when the gauge
variables are ordered (large $K$), the order-parameter field is the
gauge-dependent $N$-vector field, whose critical behavior can only be
probed by using a stochastic gauge fixing that reduces the gauge
freedom.
\end{abstract}

\maketitle

\section{Introduction}
\label{intro}

Gauge symmetries and Higgs phenomena are key features of theories
describing high-energy particle physics~\cite{Weinberg-book} and
collective phenomena in condensed-matter
physics~\cite{Anderson-book,Wen-book,Fradkin-book,Sachdev-19}. In both
contexts, it is crucial to have a solid understanding of the interplay
between global and gauge symmetries, and, in particular, of the role
that local gauge symmetries play in determining the phase structure of
a model, the nature of its different phases and of its quantum and
thermal transitions. Several lattice Abelian and non-Abelian gauge
models have been considered, with the purpose of identifying the
possible universality classes of the continuous transitions, see,
e.g.,
Refs.~\cite{Wegner-71,HLM-74,BDI-74,FS-79,Kogut-79,DH-81,BF-83,FM-83,
  BN-87, RS-90,MS-90,HL-91,Wen-91,LRT-93,BFLLW-96,HT-96,SF-00,HS-00,
  MSF-01,SSS-02, SM-02, KNS-02,MHS-02, SSSNH-02,Kitaev-03,SSNHS-03,
  NRR-03, SBSVF-04, MV-04, SSS-04, Nussinov-05,TIM-05, WBJSS-05,
  CIS-06, Sandvik-07, CAP-08, KS-08, VDS-09,ODHIM-09, GS-10, TKPS-10,
  GHMS-11,DKOSV-11,IMH-12, PDA-13, BMK-13, HBBS-13, BS-13, NCSOS-15,
  SP-15, KNNSWS-15, SWHSL-16,WS-16,Sachdev-16,WNMXS-17, FH-17, PTV-17,
  PTV-18, GASVW-18, IZMHS-19, Sachdev-19,BPV-19,SSST-19,ZZ-19,
  BPV-20-on,SPSS-20,BPV-21,WB-21,BFPV-21-mpsun,SSN-21,BPV-21-coAH,
  BPV-22,BPV-22-z2h,BPV-22-dis,BP-23,BPV-23-chgf,Senthil-23,SZJSM-23,
  BPV-23-mppo,BPV-24-ncAH,BPV-24-decQ2,BPV-24-strcou} for a partial
selection of references.  They provide examples of topological
transitions, which are driven by extended charged excitations with no
local order parameter, or by a nontrivial interplay between long-range
scalar fluctuations and nonlocal topological gauge modes.

In this paper we discuss the phase diagram and critical behavior of
three-dimensional (3D) lattice ${\mathbb Z}_2$-gauge $N$-vector
models, obtained by minimally coupling $N$-component real variables
with ${\mathbb Z}_2$-gauge variables.  They are interesting
paradigmatic models with different phases characterized by the
spontaneous breaking of the global O($N$) symmetry and by the
different topological properties of the ${\mathbb Z}_2$-gauge
correlations, see, e.g., Refs.~\cite{FS-79,Sachdev-19}. Moreover, they
are relevant for transitions in nematic liquid crystal, see, e.g.,
Refs.~\cite{LRT-93,KNNSWS-15}, and for systems with fractionalized
quantum numbers, see, e.g., Ref.~\cite{SSS-02,SM-02}.

\begin{figure}[tbp]
\includegraphics[width=0.9\columnwidth, clip]{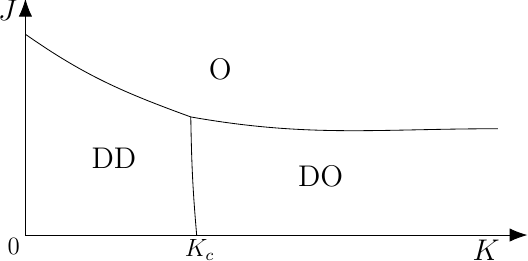}
\caption{Sketch of the phase diagram of the 3D ${\mathbb Z}_2$-gauge
  $N$-vector model with $N\ge 2$ in the space of the Hamiltonian
  parameters $K$ and $J$, cf. Eq.~(\ref{ham}), where $K$ is the
  inverse gauge coupling and $J$ is the spin hopping parameter.  There
  are two spin-disordered phases for small $J$: a small-$K$ phase, in
  which both spin and ${\mathbb Z}_2$-gauge variables are disordered
  (we indicate it by DD), and a large-$K$ phase, in which the
  ${\mathbb Z}_2$-gauge variables order (we indicate it by DO).  For
  large $J$ there is a single phase, in which both spins and gauge
  variables are ordered (we indicate it by O).  }
\label{phadiaN}
\end{figure}

The phase diagram of the 3D ${\mathbb Z}_2$-gauge $N$-vector model
presents three phases distinguished by the order/disorder of the spin
correlations and the order/disorder of the ${\mathbb Z}_2$-gauge
correlations. A sketch of the phase diagram for $N\ge 2$ is shown in
Fig.~\ref{phadiaN}.  There are two spin-disordered phases separated by
a topological ${\mathbb Z}_2$-gauge transition, and one spin-ordered
phase with topologically trivial gauge correlations.  These phases are
separated by three transition lines, whose nature crucially depends on
$N$, with the exception of the purely topological transition between
the spin-disordered phases, which belongs to the ${\mathbb Z}_2$-gauge
universality class~\cite{Wegner-71,FS-79,Sachdev-19} for any $N$
(obviously the presence of first order transitions cannot be excluded
by universality arguments).  For $N\ge 3$ the DD-O transitions are
expected to be first order, while along the DO-O transition line the
system undergoes continuous transitions belonging to the O($N$) vector
universality class. Note, however, that along the DO-O transition line
there are apparently no critical vector correlations, because of the
${\mathbb Z}_2$-gauge invariance. To identify a vector critical field,
it is necessary to reduce the gauge freedom, by performing an
appropriate gauge fixing. We use the stochastic gauge fixing outlined
in Ref.~\cite{BPV-24-gaufix}.

Unlike models with $N\ge 3$, the model with $N=2$ can develop
continuous transitions along all three transition lines.  In
particular, the DD-O and DO-O continuous transitions (see
Fig.~\ref{phadiaN}) between the spin-disordered phases and the
spin-ordered one are both expected to belong to the $XY$ universality
class.  However, this does not imply that the relevant critical modes
are the same. Indeed, as we shall see, the correlations of the
gauge-invariant operators have a different critical behavior along the
DD-O and DO-O transition lines.

To numerically check the theoretical predictions, we report Monte
Carlo (MC) simulations of the $N=2$ model in different regions of its
phase diagram. A finite-size scaling (FSS) analysis of the numerical
data confirms the general picture.  The system undergoes continuous
transitions along all three transition lines, except possibly
sufficiently close to the meeting point of the transition lines where
the transitions may turn into first-order ones.  Along the DD-O and
DO-O transition lines continuous transitions belong to the $XY$
universality class. However, while the order parameter for the DD-O
transitions is a gauge-invariant variable, the order parameter
along the DO-O transition line can be identified with the
non-gauge-invariant spin variable, after an appropriate gauge fixing
procedure (without fixing the gauge, the correlation functions of the
vector variables trivially vanish).

The paper is organized as follows. In Sec.~\ref{model} we introduce
the 3D ${\mathbb Z}_2$-gauge $N$-vector models. The phase diagram and
nature of the transition lines for $N\ge 2$ are discussed in
Sec.~\ref{phasediag}. In Sec.~\ref{numres} we present our numerical
results for $N=2$. In Sec.~\ref{multinat} we focus on the transitions
at the meeting point of the three transition lines separating the
different phases.  In Sec.~\ref{gaufix} we present results obtained by
the stochastic gauge fixing put forward in Ref.~\cite{BPV-24-gaufix},
which allows us to observe critical vector correlations along the DO-O
line.  Finally, in Sec.~\ref{conclu} we summarize and draw our
conclusions.

\section{The ${\mathbb Z}_2$-gauge $N$-vector models}
\label{model}

We consider lattice $N$-vector models with local ${\mathbb Z}_2$ gauge
invariance, defined on a 3D cubic lattice of linear size $L$ with
periodic boundary conditions. The system variables are unit-length
$N$-component real vectors ${\bm s}_{\bm x}$ (i.~e., ${\bm s}_{\bm
  x}\in\mathbb{R}^N$ and ${\bm s}_{\bm x} \cdot {\bm s}_{\bm x}=1$)
defined on the lattice sites, and ${\mathbb Z}_2$ spins $\sigma_{{\bm
    x},\mu}=\pm 1$ defined on the bonds ($\sigma_{{\bm x},\mu}$ is
associated with the bond starting from site ${\bm x}$ in the positive
$\mu$ direction, $\mu=1,2,3$).  The partition function reads
\begin{equation}
Z = \sum_{\{{\bm s},\sigma\}} e^{-H(J,K)/T}, 
  \label{partfuncmodel}
\end{equation}
where $H(J,K)$ is the lattice Hamiltonian defined by
\begin{eqnarray}
 H(J,K) = H_{\bm s}(J) + H_\sigma(K), \label{ham}
\end{eqnarray}
where
\begin{eqnarray}
&& H_{\bm s}(J) = - J N \sum_{{\bm x},\mu} \sigma_{{\bm x},\mu} \, {\bm s}_{\bm
    x} \cdot {\bm s}_{{\bm x}+\hat{\mu}}, \label{hs}\\
&& H_\sigma(K) =  - K \sum_{{\bm
      x},\mu>\nu}
   \sigma_{{\bm
      x},\mu} \,\sigma_{{\bm x}+\hat{\mu},\nu} \,\sigma_{{\bm
      x}+\hat{\nu},\mu} \,\sigma_{{\bm x},\nu}.\qquad
\label{hsi}
\end{eqnarray}
By measuring energies in units of the temperature $T$, we can formally
set $T=1$ in Eq.~\eqref{partfuncmodel}. The Hamiltonian (\ref{ham}) is
invariant under global O($N$) transformations acting on the spin
variables ${\bm s}_{\bm x}$, and under local ${\mathbb Z}_2$ gauge
transformations, ${\bm s}_{\bm x}\to w_{\bm x} {\bm s}_{\bm x}$ and
$\sigma_{{\bm x},\nu}\to w_{\bm x} \sigma_{{\bm x},\nu}w_{{\bm
    x}+\hat{\nu}}$ with $w_{\bm x}=\pm 1$. For $N=1$ the spin
variables take the integer values $s_{\bm x}=\pm 1$, and the model
corresponds to the so-called ${\mathbb Z}_2$ gauge Higgs
model~\cite{Wegner-71,FS-79,Kogut-79}.

The critical behavior at the phase transitions can be determined by
analyzing the FSS behavior of gauge-invariant correlation functions.
For this purpose, for $N\ge 2$, we consider the spin-two bilinear
operator
\begin{eqnarray}
  Q^{ab}_{\bm x} = s_{\bm x}^a s_{\bm x}^b - {1\over N}\delta^{ab},
  \label{qab}
  \end{eqnarray}
and its correlation function
\begin{eqnarray}
  G({\bm x},{\bm y}) = \langle {\rm Tr} \,Q_{\bm x} Q_{\bm y} \rangle.
  \label{gqdef}
\end{eqnarray}
The corresponding susceptibility $\chi$ and second-moment correlation
length $\xi$ are defined as
\begin{eqnarray}
  \chi \equiv  \widetilde{G}({\bm 0}),\quad
  \xi^2 \equiv {1\over 4 \sin^2 (\pi/L)}
     {\widetilde{G}({\bm 0})
       - \widetilde{G}({\bm p}_m)\over \widetilde{G}({\bm p}_m)},
\label{xidefpb}
\end{eqnarray}
where $\widetilde{G}({\bm p})=\sum_{{\bm x}} e^{i{\bm p}\cdot {\bm x}}
G({\bm x})$ and ${\bm p}_m=(2\pi/L,0,0)$.  We also consider
renormalization-group (RG) invariant quantities, whose scaling
behavior does not depend on any nonuniversal normalization, such as
the ratio
\begin{equation}
 R \equiv \xi/L, 
 \label{defR}
\end{equation}
and the Binder parameter defined as
\begin{equation}
  U = {\langle m_{2}^2\rangle\over \langle m_{2} \rangle^2}, \qquad
  m_{2} = {1\over L^d} \sum_{{\bm x},{\bm y}} {\rm Tr} \,Q_{\bm x} Q_{\bm y}.
  \label{defU}
\end{equation}

\section{The phase diagram}
\label{phasediag}

\begin{figure}[tbp]
\includegraphics[width=0.95\columnwidth, clip]{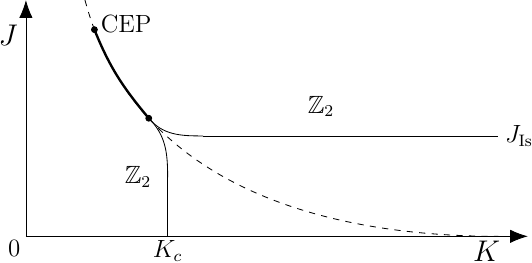}
\caption{Sketch of the phase diagram of the 3D ${\mathbb Z}_2$ gauge
  Higgs model, see, e.g., Refs.~\cite{BPV-22,SSN-21,TKPS-10}. The
  dashed line is the self-dual line, the thick line is a finite
  stretch of the self-dual line corresponding to first-order
  transitions.  The two lines labelled ``${\mathbb Z}_2$" are related
  by duality, and correspond to Ising continuous transitions.  They
  end at $[J = J_{\rm Is} \approx 0.221655, K=\infty]$ and at $[J =0,K
    = K_{{\mathbb Z}_2} \approx 0.761413]$.  The three lines meet at a
  multicritical point ~\cite{TKPS-10,SSN-21,BPV-22,OKGR-23,XPK-24} at
  $[K_\star=0.7525(1),J_\star\approx 0.22578(5)]$. The corresponding
  multicritical behavior is controlled by the multicritical $XY$ fixed
  point~\cite{BPV-22,BPV-24-com,XPK-24}. The second endpoint (CEP) of
  the first-order transition line, at $[K\approx 0.688,J\approx
    0.258]$, is expected to be an Ising critical endpoint.}
\label{phadiaz2}
\end{figure}

The phase diagram of the ${\mathbb Z}_2$ Higgs model, corresponding to
the model (\ref{ham}) with $N=1$, has already been thoroughly
investigated, see, e.g., Refs.~\cite{BPV-22,SSN-21,TKPS-10}.  Its
phase diagram is reported in Fig.~\ref{phadiaz2}.  In the following we
focus on the multicomponent cases, $N\ge 2$.  To understand their phase
diagram, we first consider some limiting cases, corresponding to
simpler models whose thermodynamic behavior is already known. Their
transition points are then expected to be the starting points of
transition lines, which separate the different phases of the model
with Hamiltonian (\ref{ham}).

\subsection{The transition line starting from $J=0$}
\label{J0trline}

For $J=0$ the model reduces to the ${\mathbb Z}_2$ gauge
model~\cite{Wegner-71} for any $N$. Therefore, there is a continuous
topological phase transition along the line $J=0$, which is that of
the ${\mathbb Z}_2$ gauge model, at~\cite{BPV-20-hcAH,FXL-18}
$K_{{\mathbb Z}_2}=0.761413292(11)$, separating a confined phase at
small $K$ from a deconfined phase at large $K$. This critical point is
expected to be the starting point of a transition line, which
separates two phases with different topological ${\mathbb Z}_2$ gauge
properties~\cite{FS-79}: the gauge modes are disordered for small $K$,
ordered in the oppostite case (see, e.g., Ref.~\cite{Sachdev-19}).
Both phases are disordered with respect to the spin variables.

Since the spin variables are not critical for sufficiently small
values of $J$, they can be integrated out. At leading order in $J$,
one obtains again the ${\mathbb Z}_2$ gauge
model~\cite{Wegner-71,FS-79,Kogut-79,LRT-93}, with a renormalized
gauge coupling $K$, i.e., $K \to K(J) = K + N J^4$.  This result
indicates that the transition line starting from $(J=0,K=K_{{\mathbb
    Z}_2})$ bends toward small values of $K$, as
\begin{equation}
  K_c(J) = K_{{\mathbb Z}_2} - N J^4 + O(J^6).
  \label{kcj}
\end{equation}
The existence of such topological transition line should be limited to
the region where the spin variables ${\bm s}_{\bm x}$ are disordered,
therefore, for sufficiently small values of $J$.

We also mention that no phase transitions are expected in the opposite
limit $J\to\infty$, where the spin and gauge variables order. In this
limit, modulo gauge transformations, we can set ${\bm s}_{x} = {\bm
  e}$ and $\sigma_{{\bm x},\mu} = 1$, where ${\bm e}$ is a unit
vector.

\subsection{The transition line starting from $K=0$}
\label{K0trline}

For $K=0$, the ${\mathbb Z}_2$ gauge variables can be easily
integrated out, obtaining a lattice formulation of the so-called
$RP^{N-1}$ model, whose Hamiltonian is
\begin{eqnarray}
  H_{K=0} &=& -
  \sum_{{\bm x},\mu} \ln\left[2\cosh(J N\,{\bm s}_{\bm x} \cdot {\bm
      s}_{{\bm x}+\hat{\mu}})\right]
  \label{rpm}\\
  &=& -\sum_{{\bm x},\mu}\left[
    \ln 2  + {J^2 N^2\over 2} |{\bm s}_{\bm
    x} \cdot {\bm s}_{{\bm x}+\hat{\mu}}|^2 + O(J^4)\right].
  \nonumber
\end{eqnarray}
Like the standard $RP^{N-1}$ model with Hamiltonian
$H_{RP}=-J'\sum_{{\bm x},\mu} |{\bm s}_{\bm x} \cdot {\bm s}_{{\bm
    x}+\hat{\mu}}|^2$, the variant model with Hamiltonian (\ref{rpm})
is expected to undergo a phase transition for any $N\ge 2$ (no phase
transitions occur at $K=0$ for $N=1$, see Fig.~\ref{phadiaz2}, because
the Hamiltonian $H_{RP}$ is trivial in this case).

Since the gauge modes are not critical, the nature of the phase
transitions in $RP^{N-1}$ models can be inferred by means of a
standard Landau-Ginzburg-Wilson (LGW) argument. We consider a field
$\Phi^{ab}$, which is a symmetric traceless matrix obtained
by coarse-graining the order parameter (\ref{qab}), and the LGW
Hamiltonian (see, e.g., Refs.~\cite{PTV-17,PTV-18})
\begin{eqnarray}
{\cal L}_{\rm LGW} &=& {\rm Tr} (\partial_\mu \Phi)^2 
+ r \,{\rm Tr} \,\Phi^2 \label{hlg}\\
&+&   w \,{\rm tr} \,\Phi^3 
+  \,u\, ({\rm Tr} \,\Phi^2)^2  + v\, {\rm Tr}\, \Phi^4.
\nonumber
\end{eqnarray}
For $N=2$ the Lagrangian (\ref{hlg}) is equivalent to that of the $XY$
vector model (in particular, the $\Phi^3$ term cancels). Thus,
continuous transitions should belong to the $XY$ universality
class~\cite{PV-02}. For larger values of $N$, the LGW approach
predicts all transitions to be of first order, because of the presence
of the $\Phi^3$ term, see, e.g., Ref.~\cite{PV-19-AH3d}.

A natural hypothesis is that a transition line starts from the
transition point at $K=0$, with the same critical behavior as for
$K=0$.  Therefore, for small values of $K$ we expect a continuous $XY$
transition line for $N=2$ and a first-order transition line for any
$N\ge 3$.

\subsection{The transition line starting from $K=\infty$}
\label{Kinftytrline}

For $K\to \infty$, the plaquettes must take their maximum value, i.e., 
\begin{equation}
\Pi_{{\bm x},\mu\nu} = \sigma_{{\bm x},\mu} \,\sigma_{{\bm
    x}+\hat{\mu},\nu} \,\sigma_{{\bm x}+\hat{\nu},\mu} \,\sigma_{{\bm
    x},\nu} = 1.
  \label{frupla}
\end{equation}
Therefore, in infinite volume, modulo gauge transformations, we can
set $\sigma_{{\bm x},\mu} = 1$. The Hamiltonian (\ref{ham}) coincides
therefore with that of the standard lattice $N$-vector model without
gauge variables. It follows that, for $K\to\infty$, the system
undergoes a continuous transition at $J_{c}(K=\infty) = J_{c,{\rm
    O}(N)}$, belonging to the O($N$) vector universality
class. Estimates of the critical point $J_{c,{\rm O}(N)}$ in
$N$-vector models can be found in
Refs.~\cite{Hasenbusch-19,DBN-05,BFMM-96,Hasenbusch-22,DPV-15,BC-97,CPRV-96}.
In particular, $J_{c,{\rm O}(2)}=0.22708234(9)$ for
$N=2$~\cite{Hasenbusch-19}, and $J_{c,{\rm O}(N)}=0.252731...$ for
$N\to\infty$~\cite{CPRV-96}.

It is again natural to conjecture the existence of a transition line
that starts from the O($N$) transition point for $K\to\infty$. Along
this line, for sufficiently large values of $K$, we expect transitions
to belong to the O($N$) universality class as for $K=\infty$.  Indeed,
if the probability that $\Pi_{{\bm x},\mu\nu} = -1$ is sufficiently
small (as expected in the large-$K$ and low-$J$ phase), the nature of
the transition should be the same as for $K=\infty$.  The stability of
the $K\to\infty$ O($N$)-vector fixed point against gauge fluctuations
is essentially related to the discrete nature of the gauge variables,
whose fluctuations are suppressed in the topologically ordered
phase. Note that, in the presence of continuous Abelian and
non-Abelian gauge symmetries, gauge interactions destabilize the
$N$-vector critical behavior observed for $K=\infty$, In this case,
even for large values of $K$, transitions along the line that ends in
the $N$-vector critical point for $K=\infty$, do not belong to the
$N$-vector universality class and have a different nature, see, e.g.,
Refs.~\cite{BPV-19,BPV-22,BPV-23-chgf,BPV-23-mppo,BPV-24-ncAH}

We remark that the prediction that the large-$K$ transitions belong to
the $N$-vector universality class is apparently in contradiction with
what one would obtain from a naive application of the LGW approach.
Indeed, in $N$-vector transitions the order parameter is the
magnetization, i.e., the spin ${\bm s}_{\bm x}$. But, the spin is not
gauge invariant, and indeed the vector correlation function $\langle
{\bm s}_{\bm x} \cdot {\bm s}_{\bm y}\rangle$ vanishes for ${\bm x}
\not= {\bm y}$.  Only gauge-invariant operators are critical, the
simplest one being the spin-two operator introduced in
Eq.~\eqref{qab}. Thus, in a naive application of the LGW approach, one
would reason as in Sec.~\ref{K0trline}, obtaining the LGW Lagrangian
(\ref{hlg}), and therefore predicting the transition lines DD-O and
DO-O to have the same nature. These conclusions contradict the
arguments given above for the large-$K$ transition line.

In the following, we will provide robust evidence that the large-$K$
transitions belong to the O($N$) vector universality class. This
implies that the naive LGW argument is incorrect when applied to the
DO-O transitions. Indeed, along this line, the order parameter turns
out to be a vector field, like the standard $N$-vector models,
although such vector order parameter cannot be directly identified in
the ${\mathbb Z}_2$-gauge $N$-vector models, because of the gauge
invariance. It emerges only once an appropriate gauge fixing is
introduced, as discussed in Ref.~\cite{BPV-24-gaufix}.  Note that an
appropriate gauge fixing is needed not only for finite values of $K$
but also for $K=\infty$. Indeed, we obtain the $N$-vector Hamiltonian
only if we fix the gauge so that $\sigma_{{\bm x},\mu} = 1$ on all
links.

It is worth noting that, for $N=2$, the LGW model with Lagrangian
(\ref{hlg}) predicts an $XY$ critical behavior as the corresponding
$N$-vector LGW theory. However, in the first case $Q_{\bm x}^{ab}$
behaves as a two-component vector field, while in the second case
$Q_{\bm x}^{ab}$ is a spin-two operator. Thus, the critical behavior
of $Q_{\bm x}^{ab}$ allows us to distinguish which is the appropriate
LGW description of the transition. For $N\ge 3$ the LGW predictions
are different.  In particular, the theory with Lagrangian (\ref{hlg})
predicts first-order transitions because of the cubic term. Even
admitting the possibility that the cubic term somehow vanishes, as for
antiferromagnetic $RP^{N-1}$ models, the critical behavior would be
different from the O($N$)-vector one (see Ref.~\cite{PTV-18} for a RG
analysis of the theory with Lagrangian (\ref{hlg}) and $w=0$).

\subsection{The $J$-$K$ phase diagram}
\label{phadia}

To draw the phase diagram in the $J$-$K$ parameter space of the
$N$-component model, we make the natural hypothesis that the
transitions identified along the lines $K=0$, $K=\infty$, and $J=0$
are the starting points of three transition lines that meet in a
single point, as sketched in Fig.~\ref{phadiaN}. These transition
lines are the boundaries of three different phases. For small values
of $J$ we expect two phases in which the spin variables are
disordered. For small $K$ also the gauge degrees of freedom are
disordered---we name this phase disordered-disordered (DD) phase. For
large $K$, instead, gauge variables are topologically ordered---this
is the disordered-ordered (DO) phase.  For large $J$ there is instead
a single phase, in which both spin and gauge variables are
ordered---we name it ordered (O) phase.  In this phase the bilinear
spin-two operator $Q_{\bm x}^{ab}$ condenses.

As already discussed, for sufficiently small $K$, the DD-O transitions
should have the same nature as the $RP^{N-1}$ transition along the
$K=0$ line. The corresponding critical behavior id therefore
controlled by the LGW theory (\ref{hlg}).  Instead, for sufficiently
large values of $K$ along the DO-O transition line, we expect
transitions to belong to the O($N$) vector universality class.  Spin
variables should not play any role along the DD-DO transition, which
should have the same topological nature as the transition in the pure
${\mathbb Z}_2$ gauge theory.

\subsection{Meeting point of the transition lines}
\label{meetpoint}

The three transition lines are expected to eventually meet at one
point $[K_\star,J_\star]$ of the phase diagram. Eq.~(\ref{kcj})
suggests that $K_\star\lesssim K_{{\mathbb Z}_2}\approx 0.761$ at
least for small $N$. Indeed, if we assume that the DD-O line $J =
J_c(K)$ is weakly dependent on $K$, as it occurs in the ${\mathbb
  Z}_2$ Higgs model whose phase diagram in shown in
Fig.~\ref{phadiaz2}, the correction term in Eq.~(\ref{kcj}) is small
(of the order of $N J_{c,{\rm O}(N)}^4$, with $J_{c,{\rm O}(N)} \approx 0.2$). .
This is consistent with the results reported in Sec.~\ref{multinat},
where we argue that $K_\star \approx 0.75$ for $N=2$.

At the meeting point the system may develop a multicritical behavior
if some of the transition lines are continuous at the meeting point,
as it happens in the ${\mathbb Z}_2$ gauge Higgs model (see
Fig.~\ref{phadiaz2}). Alternatively, if all transition lines are of
first order, the meeting point corresponds to a first-order
transition.  We shall return to this point in Sec.~\ref{multinat}.

\subsection{Phase behavior for  $N=2$}
\label{n2case}

We now focus on the two-component model, which is particularly
interesting because it is the only case in which continuous
transitions may occur along all three transition lines.  In
particular, on the basis of the above discussion, the continuous
transitions along the DD-O and DO-O lines are expected to both belong
to the $XY$ universality class. However, as discussed in
Sec.~\ref{Kinftytrline}, the nature of the transitions along the two
lines is not the same and this shows up in the different critical
behavior of the operator $Q^{ab}_{\bm x}$ defined in Eq.~(\ref{qab}).

To characterize continuous transitions along the DD-O and DO-O lines,
we fix $K$ and vary the parameter $J$. Close to the transition, the
correlation function $G({\bm x},{\bm y})$ defined in Eq.~(\ref{gqdef})
is expected to show the asymptotic FSS behavior (we assume that the
boundary conditions preserve translation invariance)
\begin{eqnarray}
  &&G({\bm x}_1,{\bm x}_2,J,L) \approx L^{-2Y_Q}
  [{\cal G}({\bm X},W) + O(L^{-\omega})],\quad
  \label{g2sca} \\
  && {\bm X} = ({\bm x}_1-{\bm x}_2)/L,\quad W = (J - J_{c})
  L^{1/\nu}, \label{Wdef}
\end{eqnarray}
where $Y_Q$ is the RG dimension of the operator $Q_{\bm
  x}^{ab}$. Since both DD-O and DO-O transitions belong to the $XY$
universality class, we have $\nu=\nu_{XY}=0.6717(1)$ and
$\omega=\omega_{XY}=0.789(4)$~\cite{GZ-98,KP-17,CHPV-06,Hasenbusch-19,
  CLLPSSV-20}. However, the RG dimension $Y_Q$ differs along the DD-O
and DO-O transition lines.

Along the DD-O line, an effective description is provided by the LGW
model with Lagrangian (\ref{hlg}). In this case the coarse-grained
field $\Phi^{ab}$ is equivalent to an O(2) vector field.  This implies
that the RG dimension $Y_Q$ of $Q_{\bm x}^{ab}$ coincides with the RG
dimension $Y_{V,XY}$ of the vector field in the standard $XY$ model.
Thus, at continuous transitions along the DD-O line, we have
\begin{eqnarray}
  Y_Q = Y_{V,XY} = {d-2+\eta_{XY}\over 2}=0.519088(22)
  \label{Yspin1}
\end{eqnarray}
where we used the
precise estimate~\cite{CLLPSSV-20} $\eta_{XY}=0.038176(44)$.

On the other hand, the continuous transitions along the DO-O
transition line are expected to belong to the $N$-vector universality
class.  Therefore, the bilinear operator $Q$ corresponds to the tensor
spin-two operator in the $XY$ model, whose RG dimension $Y_{T,XY}$ has
been computed by various methods, see
Refs.~\cite{CLLPSSV-20,HV-11,CPV-03,CPV-02}. Therefore, we expect
\begin{eqnarray}
  Y_Q = Y_{T,XY} = 1.23629(11)
  \label{Yspin2}
\end{eqnarray}
at the continuous transitions along the DO-O line.

As a consequence of the above results, the susceptibility defined in
Eq.~(\ref{xidefpb}) has a substantially different dependence on the
size of the system at critical points along the two lines, a
difference that can be easily detected in FSS analyses. Since $\chi$
behaves as
\begin{equation}
  \chi \approx  L^{d-2Y_Q} {\cal A}(W) 
  \label{chisca} 
\end{equation}
in the FSS limit, we obtain
\begin{eqnarray}
\chi \sim L^{\kappa_v},\quad \kappa_v = 3 - 2 Y_{V,XY} = 1.96182(2),
\label{chiddo}
\end{eqnarray}
along the DD-O transition line, and
\begin{eqnarray}
  \chi \sim L^{\kappa_t},\quad \kappa_t = 3-2 Y_{T,XY} = 0.5274(2),
       \label{chidoo}
\end{eqnarray}
along the DO-O transition line. Also scaling corrections are expected
to be different in the two cases.  Along the DD-O transition line,
corrections are expected to scale as $L^{-\omega_{XY}}$,
where~\cite{Hasenbusch-19} $\omega_{XY} = 0.789(4)$ is associated with
the leading irrelevant operator.  On the other hand, along the DO-O
line the dominant scaling corrections to Eq.~(\ref{chidoo}) are due to
the background analytic term~\cite{PV-02}. Thus, corrections scale as
$L^{-\kappa_t}$, with $\kappa_t\approx 0.527<\omega_{XY}\approx
0.789$.  As we have discussed in Sec.~\ref{Kinftytrline}, along the
DO-O line there are also emerging vector critical modes, that shows up
only if a proper gauge fixing is introduced. They are discussed in
Sec.~\ref{gaufix}.

\section{Numerical results for the
  ${\mathbb Z}_2$-gauge N=2 vector model}
\label{numres}

To investigate the nature of the transition lines of the ${\mathbb
  Z}_2$-gauge model for $N=2$, we have performed MC simulations close
to the transition lines DD-O and DO-O, on lattices of size $L\le 40$.
Simulations have been performed by using a standard Metropolis update
for the gauge variables $\sigma_{{\bm x},\mu}$, and a combination of
Metropolis and microcanonical updates (in the ratio 1:5) for the
variables ${\bm s}_{\bm x}$. In all cases we performed MC runs of
about $5\times 10^6$ sweeps (a sweep corresponds to a complete update
of all lattice gauge and spin variables).  Simulations took a total
CPU time of roughly $1.5\times 10^5$ core-hours.

We have studied the critical behavior fixing $K$ and varying $J$. In
Table~\ref{tabres}, we report the values of $K$ considered, the
transition points $J_c(K)$, and some information on the critical
behavior.

\begin{table}
\begin{tabular}{cclcc}
\hline\hline $K$ & $\quad$ & $\;\;\;J_c(K)$ & type & $Y_Q$ \\
\hline
0   && 0.79305(7) & $XY$ & $Y_{V,XY}$ \\
0.5 && 0.37118(2) & $XY$ & $Y_{V,XY}$ \\
0.7 && 0.2520(3)  & 1$^{\rm st}$-order & \\
0.8 && 0.229(1)   & $XY$ & $Y_{T,XY}$ \\
1   && 0.22729(3) & $XY$ & $Y_{T,XY}$ \\
$\infty$  && 0.22708234(9) & $XY$ & $Y_{T,XY}$ \\
\hline\hline
\end{tabular}
\caption{Results obtained for $N=2$, varying $J$ across the DD-O and
  DO-O transition lines for fixed values of $K$.  We report the
  critical point $J_c(K)$ (for $K\to\infty$ we quote the estimate of
  the $XY$ critical point reported in Ref.~\cite{Hasenbusch-19}), the
  transition type, and, if the transition is continuous, the value of
  the RG dimension $Y_Q$ of the gauge-invariant operator $Q_{\bm
    x}^{ab}$ defined in Eq.~(\ref{qab}).}
\label{tabres}
\end{table}

\subsection{The small-$K$ DD-O transition line}
\label{ddoline}

We now report the results of the FSS analyses of the data obtained by
varying $J$ across the DD-O transition line, keeping $K$ fixed. We
considered three values of $K$, $K=0$, $K=0.5$, and $K=0.7$, which are
smaller than the value $K_\star\approx 0.75$ of the meeting point of
the three transition lines, see Sec.~\ref{meetpoint} and
Sec.~\ref{multinat}. We anticipate that the FSS analyses show that the
system undergoes continuous $XY$ transitions for $K=0$ and $K=0.5$,
and a first-order transition for $K=0.7$. Thus, the continuous $XY$
transition line starting at $K=0$ turns into a first-order line at
$K=K_{\rm fo}$ with $0.5<K_{\rm fo}<0.7$, before reaching the point where the
transition lines meet, see Sec.~\ref{multinat}.

\begin{figure}[tbp]
\includegraphics[width=0.95\columnwidth, clip]{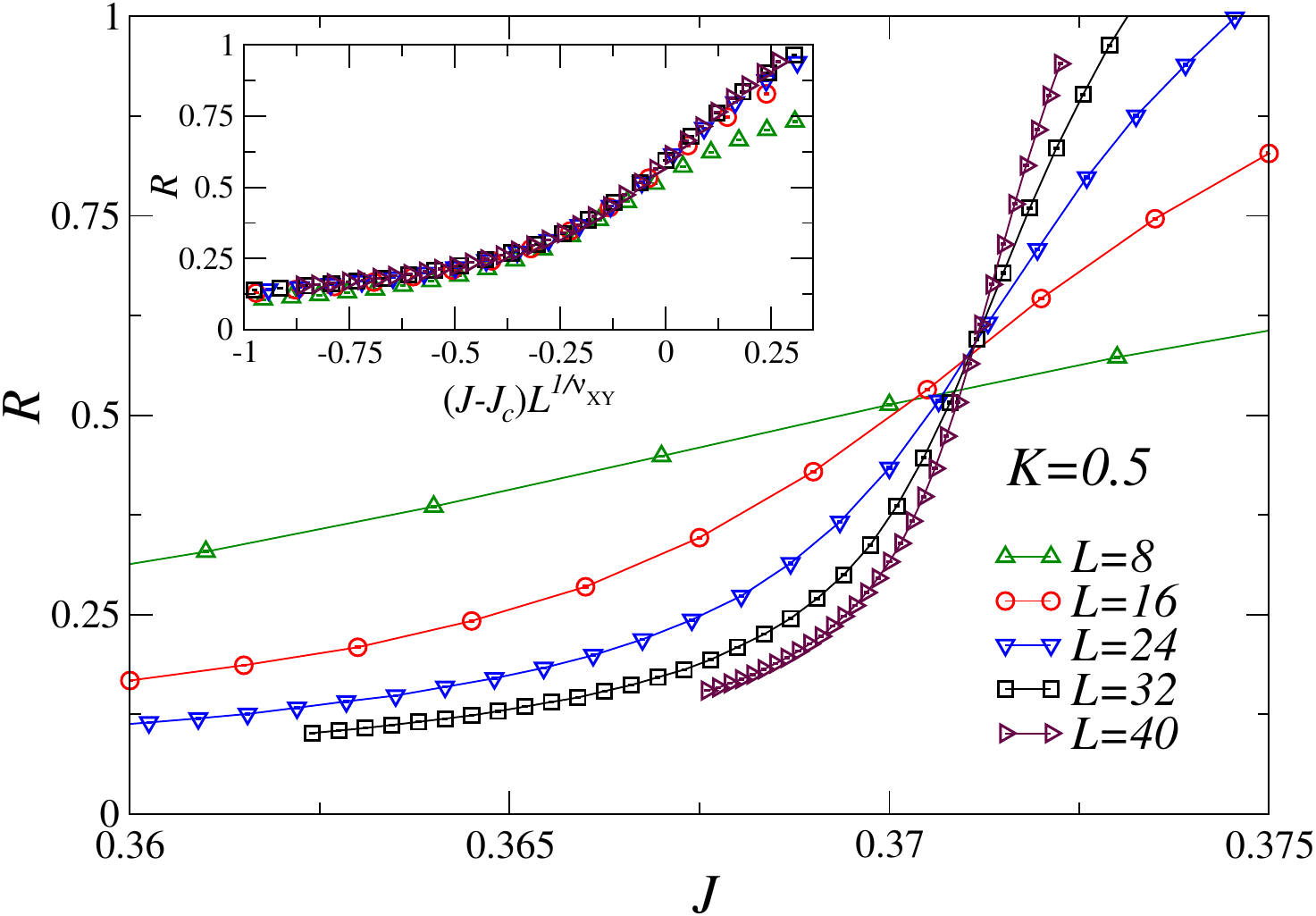}
\caption{Data of the ratio $R\equiv \xi/L$ as a function of $J$ at
  fixed $K=0.5$.  The inset shows that the data nicely collapse onto a
  single curve when $R$ is plotted $W=(J-J_c)L^{1/\nu_{XY}}$, with
  $J_c=0.37118$ and $\nu_{XY} = 0.6717$, confirming the $XY$ nature of
  the transition. }
\label{figN2_R_J_K0p5}
\end{figure}

\begin{figure}[tbp]
\includegraphics[width=0.95\columnwidth, clip]{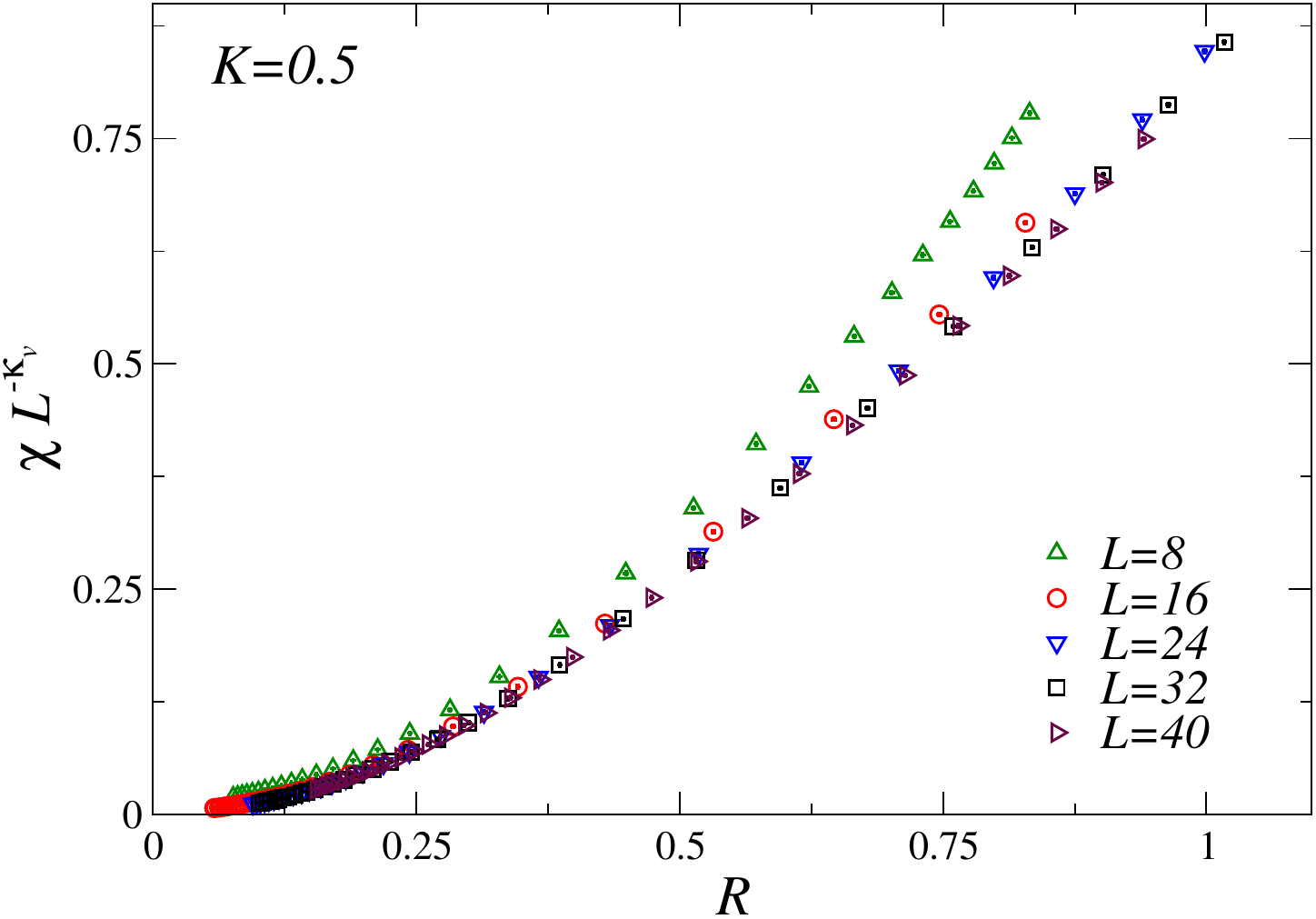}
\caption{Scaling of the susceptibility $\chi$ along the $K=0.5$ line:
  plot of $L^{-\kappa_v} \chi$ versus $R=\xi/L$, with $\kappa_v=3 -
  2Y_{V,XY} = 1.96182$. Here $Y_{V,XY}$ is the RG dimension of the
  vector field in the $XY$ universality class. The data approach an
  asymptotic scaling curve with increasing $L$, thus supporting
  relation~(\ref{chiddo}).}
\label{figN2_chi_K0p5}
\end{figure}

We first report results along the line $K=0.5$. To determine the
critical point and the order of the transition, we consider the RG
invariant quantities $R$ and $U$ defined in Eqs.~(\ref{defR}) and
(\ref{defU}).  At continuous transitions they are expected to scale as
\begin{eqnarray}
  &&R(J,L) = {\cal R}(W) + O(L^{-\omega}), \quad W = (J - J_{c})
  L^{1/\nu},\qquad \label{Rscal}\\ &&U(J,L) = {\cal U}(W) +
  O(L^{-\omega}). \label{uscal}
\end{eqnarray}
In particular, the curves obtained for different lattice sizes should cross at
the critical point, apart from scaling corrections.  In
Fig.~\ref{figN2_R_J_K0p5} we plot $R$ as a function of $J$.  The data
show a crossing point, indicating the presence of a continuous
transition at $J_c\approx 0.37$. Analogous results are obtained for
the Binder parameter.  The slopes of the data at the crossing point
are fully consistent with the length-scale exponent
\cite{Hasenbusch-19} $\nu_{XY}=0.6717(1)$ of the $XY$ universality
class.  To obtain a precise estimate of $J_c$, we have fitted the data
to Eq.~(\ref{uscal}) setting $\nu = \nu_{XY} = 0.6717$. We obtain
$J_c= 0.37118(2)$. A scaling plot is shown in the inset of
Fig.~\ref{figN2_R_J_K0p5}. We observe a very nice collapse of the
data, confirming that the transition belongs to the $XY$ universality
class.

The RG dimension $Y_Q$ of the operator $Q_{\bm x}^{ab}$ can be
estimated by fitting the susceptibility to
Eq.~(\ref{chisca}). However, from a numerical point of view, it is
more convenient to consider the FSS behavior of $\chi$ in terms of
$R\equiv \xi/L$. Indeed, since $R$ is monotonic, we can express $W$ as
a function of $R$ using Eq.~(\ref{Rscal}). Thus, we can rewrite
Eq.~(\ref{chisca}) as
\begin{equation}
  \chi(J,L) \approx L^{-(d-2 Y_Q)}\left[ \widehat{\cal A}(R) +
    O(L^{-\omega})\right].
  \label{chisca2} 
\end{equation}
The data shown in Fig.~\ref{figN2_chi_K0p5} are consistent with
Eq.~(\ref{chisca2}) using the exponent reported in Eq.~(\ref{chiddo}),
i.e., $d-2Y_Q = \kappa_v= 1.96182(2)$. Thus, the correlations of
$Q_{\bm x}^{ab}$ behave as vector correlations in the standard $XY$
model.

\begin{figure}[tbp]
\includegraphics[width=0.95\columnwidth, clip]{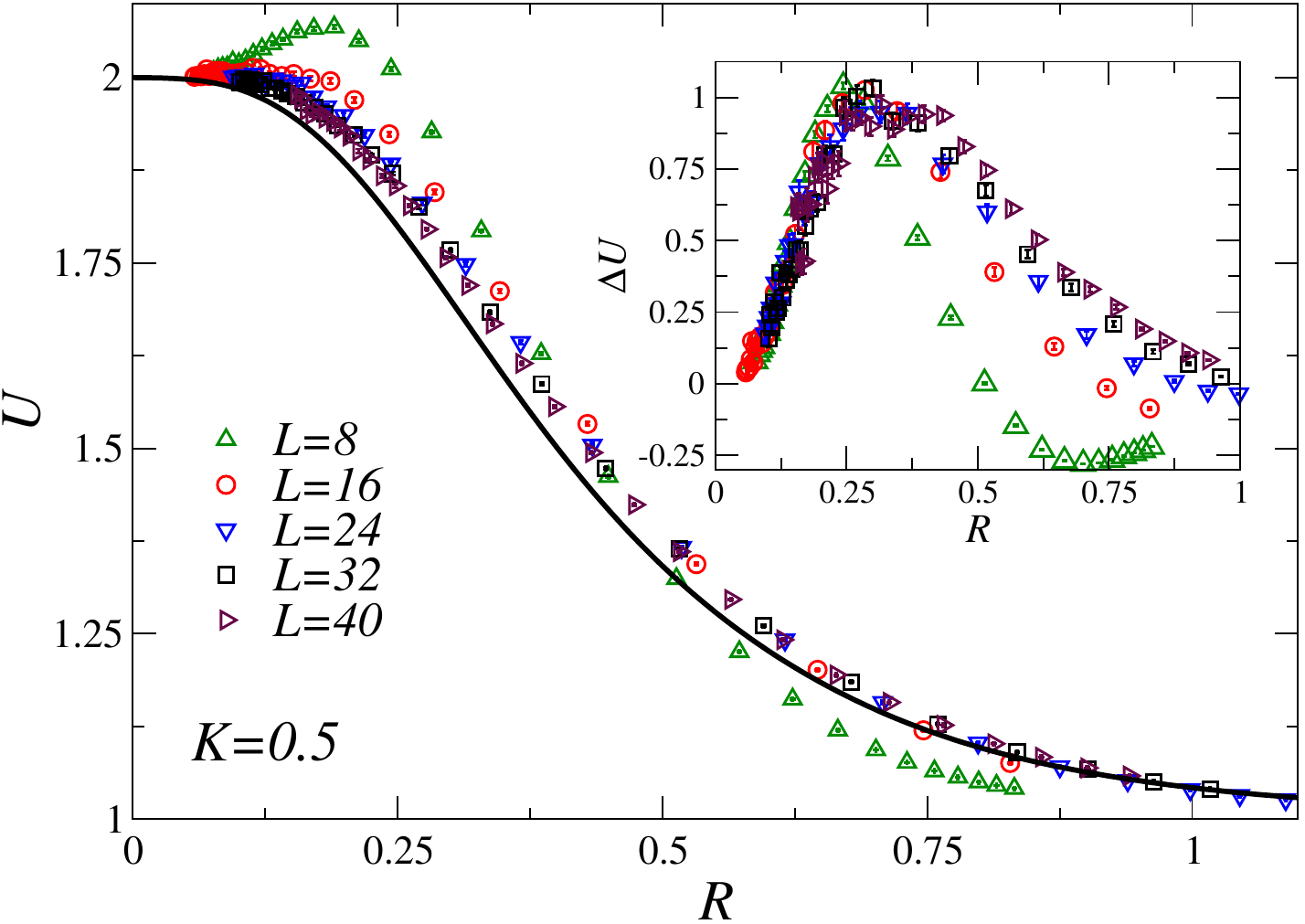}\\
\caption{The Binder parameter $U$ as a function of $R=\xi/L$ for
  $K=0.5$. The data appear to approach the universal scaling curve
  (solid line) for the $XY$ model obtained in
  Ref.~\cite{BPV-21-coAH}. The observed deviations can be explained by
  the presence of scaling corrections: In the inset we report $\Delta
  U$ defined in Eq.~(\ref{DeltaU-def}) as a function of $R$, using the
  $XY$ leading scaling-correction exponent $\omega_{XY}=0.789$. }
\label{figN2_U_K0p5}
\end{figure}

A robust check that the transition belongs to the $XY$ universality
class can be obtained by comparing the asymptotic behavior of $U$ as a
function of $R$ in the present model with the analogous data computed
in the $XY$ model. We show the data in Fig.~\ref{figN2_U_K0p5},
together with the parameterization of the $XY$ curve $U = f_{XY}(R)$
reported in Ref.~\cite{BPV-21-coAH}.  The data for the ${\mathbb
  Z}_2$-gauge $N=2$ vector model appear to approach the $XY$ scaling
curve as $L$ increases.  We observe some deviations, especially for
$L=8$ and $R\approx 0.25$, which apparently decrease as $L$ increases.
To verify that these deviations can be interpreted as scaling
corrections, we consider the quantity
\begin{equation} 
  \Delta U(J,L) = L^{\omega_{XY}} [U(J,L) - f_{XY}(R(J,L))].
\label{DeltaU-def}
\end{equation}
In the inset of Fig.~\ref{figN2_U_K0p5} we report $\Delta U$ as a
function of $R$, using the expected $XY$ correction-to-scaling exponent
$\omega_{XY}\approx 0.789$. Data fall approximately onto a single
curve as $L$ increases, providing evidence that the deviations in
Fig.~\ref{figN2_U_K0p5} are due to scaling corrections.  We remark
that FSS curves depend on boundary conditions.  Since we use here
periodic boundary conditions, we compare the data with $XY$ results
with the same boundary conditions (this is indeed the case for the
curve obtained in Ref.~\cite{BPV-21-coAH}).

We also performed simulations along the $K=0$ line, close to the
critical transition at $J_c=0.79305(7)$. The plots of $R$, $U$, and
$\chi$ are very similar to those reported in
Figs.~\ref{figN2_R_J_K0p5}, \ref{figN2_chi_K0p5}, and
\ref{figN2_U_K0p5}, so we do not report them.  Again, they confirm the
general analysis reported in Sec.~\ref{phasediag}.

\begin{figure}[tbp]
\includegraphics[width=0.95\columnwidth, clip]{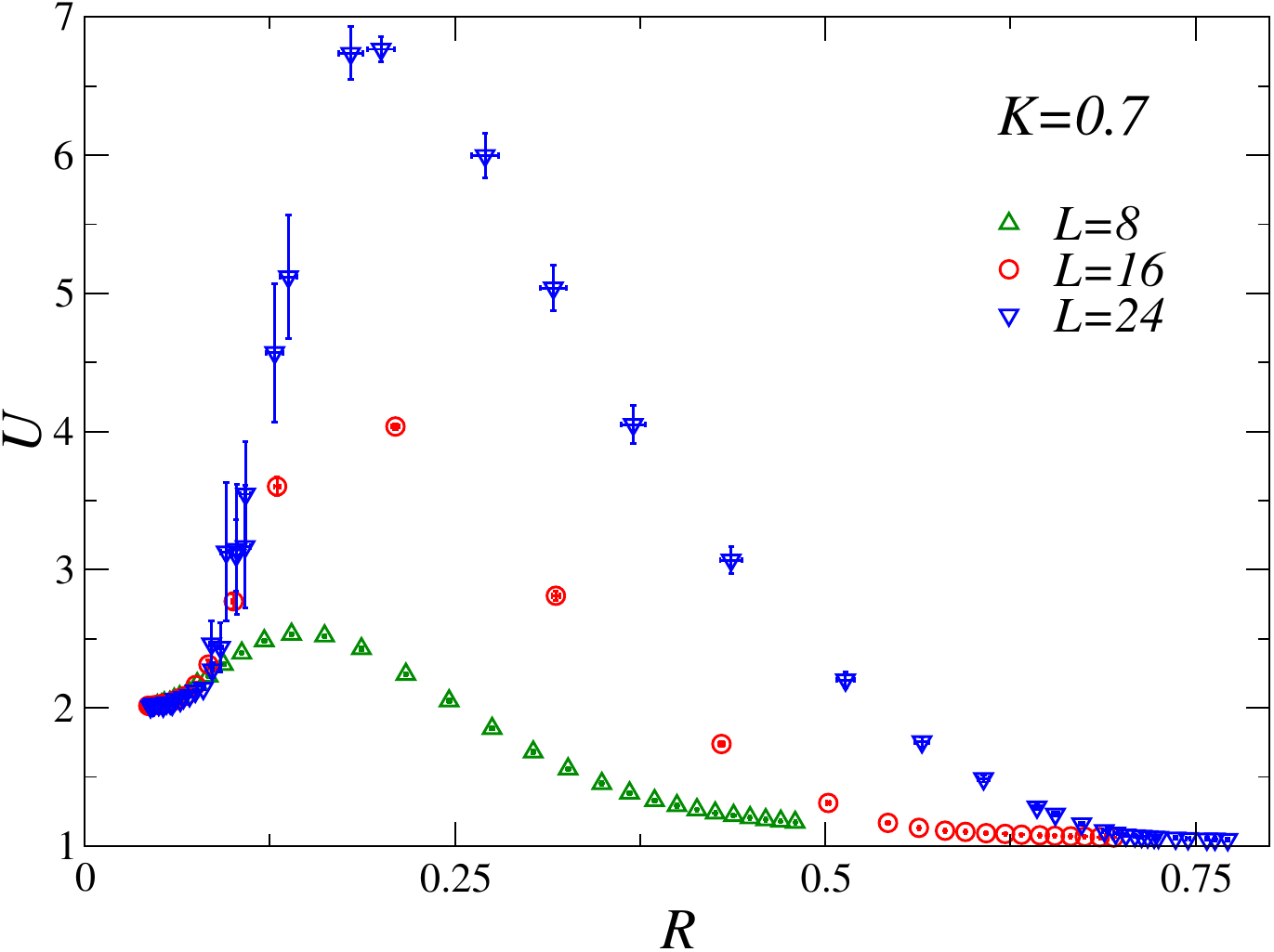}
\caption{The Binder parameter $U$ versus $R$ for $K=0.7$.  The data do
  not show scaling, and the maximum of $U$ significantly increases
  with the size, as expected at a first-order transition.}
\label{figN2_U_K0p7}
\end{figure}

Finally, we performed simulations along the line $K=0.7$.  Data
suggest a first-order transition with $J_c\approx 0.25$.  The
first-order nature of the transition can be inferred from the behavior
of the Binder parameter $U$. As shown in Fig.~\ref{figN2_U_K0p7},
fixed-$L$ data have a pronounced maximum, which significantly
increases with increasing $L$. This is usually considered as evidence
of a first-order transition, see, for example, Ref.~\cite{PV-24} and
references therein. To estimate $J_c$, we have determined the position
$J_p(L)$ of the maximum of $U$ for each size, and then we have
extrapolated the results to $J_p(L) = J_c + a/L^3$.  We obtain the
estimate $J_c=0.2520(3)$.

\subsection{The large-$K$ DO-O transition line}
\label{dooline}

We now discuss the critical behavior along the DO-O transition line.
We have performed simulations varying $J$ along the lines $K=1$ and
$K=0.8$.  These values of $K$ are larger than the value
$K_\star\approx 0.75$ corresponding to the meeting point of the three
transition lines, see Sec.~\ref{meetpoint} and Sec.~\ref{multinat}.
Therefore, for both values of $K$, we are considering DO-O
transitions.  In both cases, the FSS analyses show a continuous $XY$
transition where the operator $Q_{\bm x}^{ab}$ behaves as a spin-two
operator, in agreement with the arguments reported in
Sec.~\ref{n2case}.

\begin{figure}[tbp]
\includegraphics[width=0.95\columnwidth, clip]{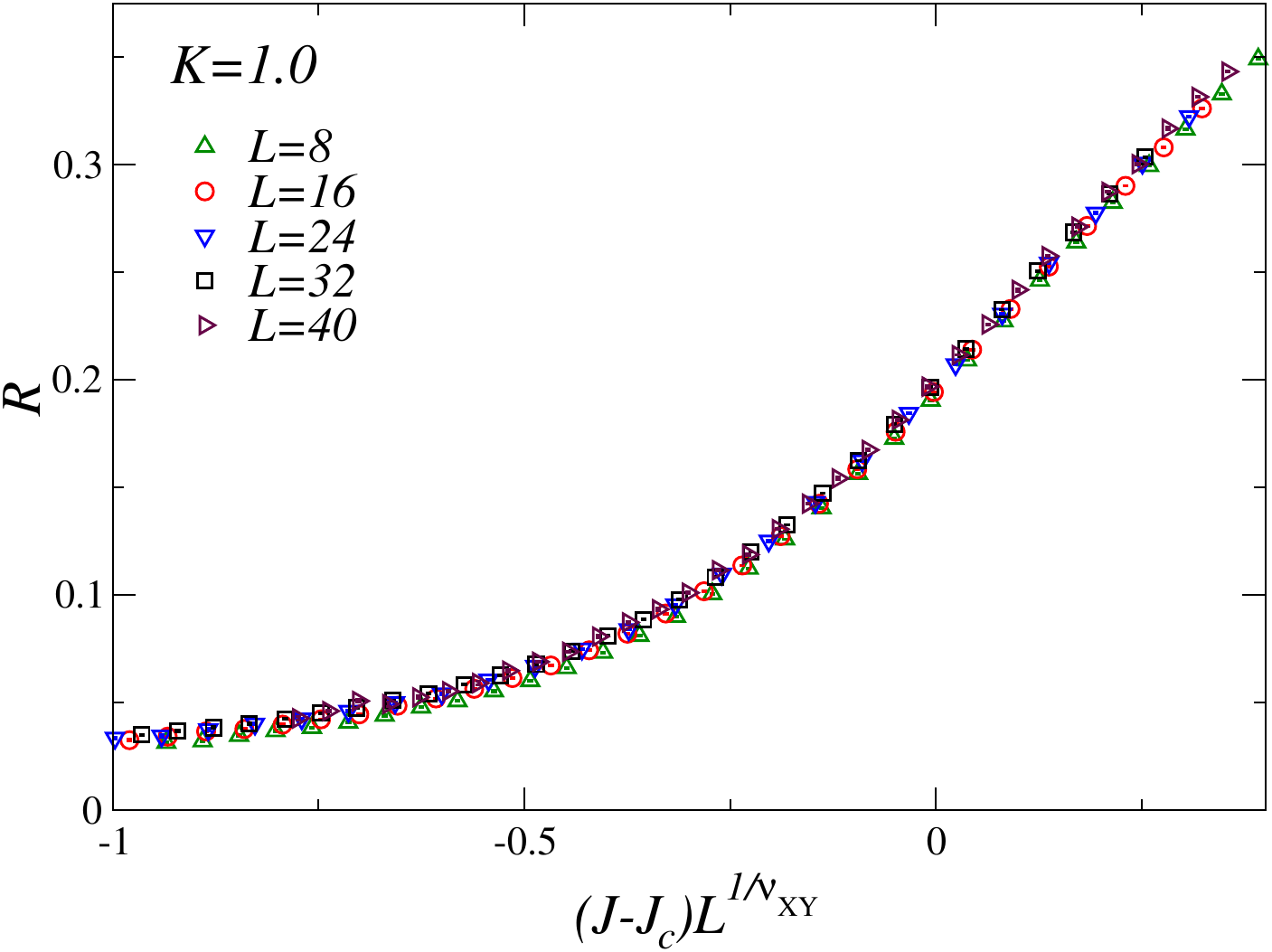}
\caption{Ratio $R\equiv \xi/L$ versus $W=(J-J_c)L^{1/\nu_{XY}}$, with
  $J_c=0.22729(3)$ and $\nu_{XY} = 0.6717$. Results for $K=1$. Data
  show an excellent collapse, confirming that the transition belongs
  to the $XY$ universality class.}
\label{figN2_R_J_K1p0}
\end{figure}

\begin{figure}[tbp]
\includegraphics[width=0.95\columnwidth, clip]{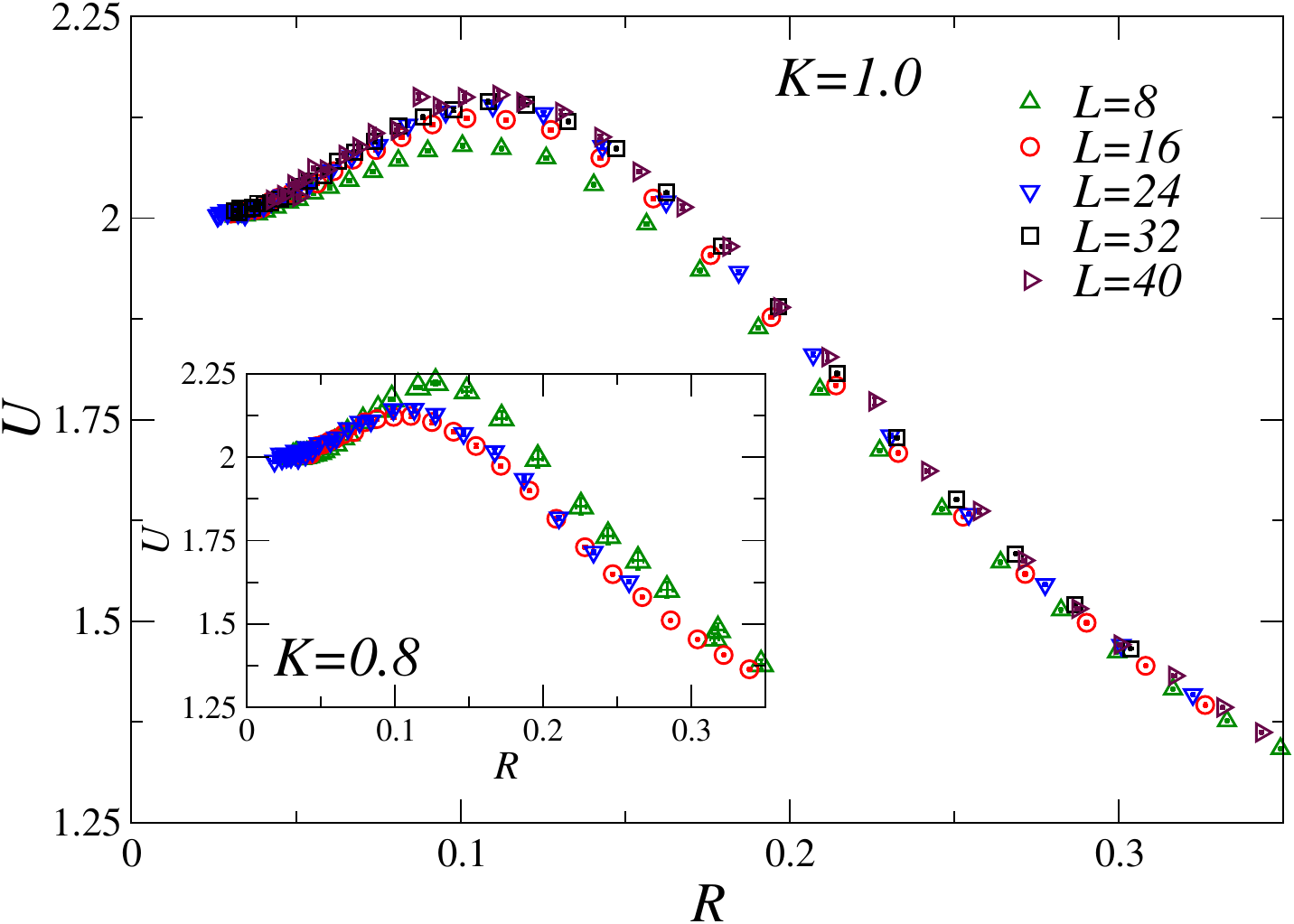}\\
\caption{The Binder cumulant $U$ as a function of $R=\xi/L$ for
  $K=1$. The data appear to collapse onto an asymptotic curve. The
  inset shows analogous data for $K=0.8$, which appear to approach the
  same asymptotic FSS curve.}
\label{figN2_U_K1p0}
\end{figure}

\begin{figure}[tbp]
\includegraphics[width=0.95\columnwidth, clip]{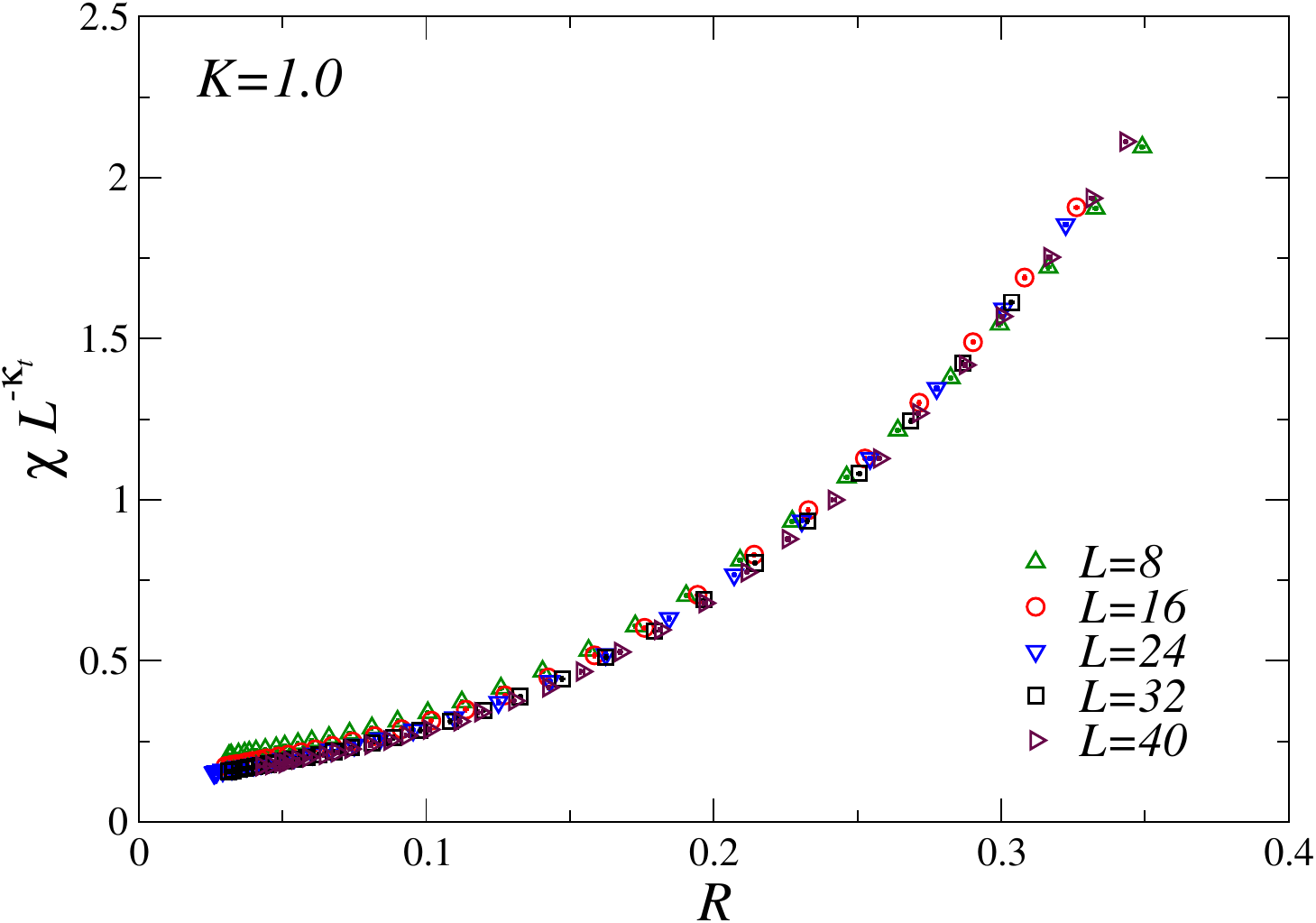}
\caption{Scaling of the susceptibility $\chi$. We plot $L^{-\kappa_t}
  \chi$ versus $R=\xi/L$, with $\kappa_t = 3-Y_{T,XY}=0.5274$.  The
  excellent collapse of the data confirms the correctness of the
  predicted exponent $\kappa_t$, see Eq.~(\ref{chidoo}).}
\label{figN2_chi_K1p0}
\end{figure}

For $K=1$, data are consistent with a continuous transition in the
$XY$ universality class.  We can accurately estimate the critical
point from the analysis of $R=\xi/L$. Fits of $R$ to Eq.~(\ref{Rscal})
using the $XY$ exponent $\nu_{XY}=0.6717$ give $J_c=0.22729(3)$.  The
corresponding scaling plot is shown in
Fig.~\ref{figN2_R_J_K1p0}. Scaling is excellent. We note that
$J_c(K=1)$ is very close to the critical value for $K=\infty$, i.e.,
\cite{Hasenbusch-19} $J_{c}(K=\infty)=0.22708234(9)$. This indicates
that the DO-O line $J=J_c(K)$ has a very weak dependence on $K$ from
$K=\infty$ to $K=1$ (as supposed in Sec.~\ref{meetpoint}).

In Fig.~\ref{figN2_U_K1p0} we report the Binder cumulant $U$ against
the ratio $R$. We observe a nice scaling that confirms the continuous
nature of the transition. Note that in this case we cannot directly
compare the results for $K=1$ with the corresponding scaling curve
computed in the $XY$ model with periodic boundary conditions. Indeed,
for $K\to\infty$ the ${\mathbb Z}_2$-gauge model with periodic
boundary conditions is equivalent to an $N$-vector model with
fluctuating boundary conditions, see the discussion in
Ref.~\cite{BPV-21-coAH}. Therefore, to perform a correct comparison
one should simulate an $XY$ model with fluctuating boundary
conditions, to determine the $XY$ curve that matches the ${\mathbb
  Z}_2$-gauge data.

In Fig.~\ref{figN2_chi_K1p0} we show a scaling plot of the
susceptibility defined in Eq.~(\ref{xidefpb}). As discussed in
Sec.~\ref{n2case}, data should scale according to Eq.~(\ref{chisca2}),
with exponent $d-2Y_Q=\kappa_t = d-2 Y_{T,XY}=0.5274(2)$.  We observe
an excellent scaling, confirming the arguments of Sec.~\ref{n2case}.

Finally, we performed simulations along the line $K=0.8$ on relatively
small lattices.  Data indicate the presence of a continuous $XY$
transition at $J_c\approx 0.229$ (the relatively low precision on
$J_c$ is due to the small lattices considered), analogous to the one
observed for $K=1$..  This is clearly demonstrated by the plots of $U$
versus $R$ shown in the inset of Fig.~\ref{figN2_U_K1p0}. The $K=0.8$
data approach the same asymptotic curve obtained for $K=1$.

\section{Transitions at the meeting point}
\label{multinat}

We now discuss the nature of the transitions close to the point
$(K_\star,J_\star)$, where the transition lines meet, see
Fig.~\ref{phadiaN}.  On the basis of the arguments reported in
Sec.~\ref{phasediag}, the DD-O transitions are of first order for any
$N\ge 3$, while they may be continuous for $N=2$. On the other hand,
for any $N$ we expect DO-O and DD-DO transitions to be continuous for
large $K$ and small $J$, respectively. Close to the meeting point,
they may be continuous or of first order, with the corresponding
presence of a tricritical point.

We wish now to estimate the position of the meeting point
$(K_\star,J_\star)$. The results reported in Table~\ref{tabres} show
that the point $[K=0.5,J = 0.37118(2)]$ belongs to the DD-O line,
while the point $[K=0.8,J \approx 0.229]$ belongs to the DO-O
line. This allows us to bound $J_\star$: $0.371 > J_\star > 0.229$. We
can then use the approximate formula (\ref{kcj}) to obtain a bound on
$K_\star$: $0.724 < K_\star < 0.756$.  This estimate of $K_\star$
allows us to conclude that the first-order transition observed at
$(K=0,7,J=0.2520(3))$ belongs to the DD-O line.  Therefore, there is a
tricritical point at $K=K_{fo}$, $0.5< K_{fo} < 0.7$ on the DD-O line,
such that DD-O transitions are continuous for $K<K_{fo}$ and of first
order in the opposite case.  The results for $K=0.7$ allow us to
improve our estimate of $J_\star$, which should belong to the interval
$[J_c(0.8),J_c(0.7)] = [0.229,0.252]$. In turn, we can use this result
to improve our estimate of $K_\star$. We obtain finally the estimates
\begin{equation}
  (K_\star\approx 0.75, \; J_\star\approx 0.24)\quad {\rm for}\;\;N=2.
  \label{kjstarn2}
\end{equation}
To verify the accuracy of these arguments, we have applied similar
arguments to the ${\mathbb Z}_2$ gauge Higgs model. For the
multicritical point, we obtain $K_\star\approx 0.75$, in good
agreement with the accurate estimate $K_\star=0.7525(1)$, see
Fig.~\ref{phadiaz2}.

\begin{figure}[tbp]
\includegraphics[width=0.95\columnwidth, clip]{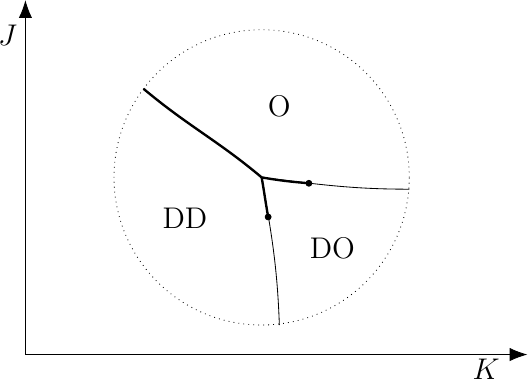}
\caption{Sketch of the phase diagram close to a first-order meeting
  point.}
\label{figMCP}
\end{figure}

Given that the DD-O transition line is of first order, the nature of
the meeting-point transition is controlled by the competition of the
$N$-vector order parameter driving the DO-O transitions (the emerging
order parameter discussed in Sec.~\ref{Kinftytrline}) and the nonlocal
order parameter driving the Ising topological transitions along the
DD-DO transition line. We are not able to define an effective model
appropriate to describe the meeting-point transition. In general, two
different behaviors are possible. In one case, the DD-DO and DO-O line
are continuous up to the meeting point, so that we obtain what is
usually called a bicritical point. Alternatively, the continuous
transitions may turn into first-order ones before the meeting point,
as it occurs along the DD-O transition lines, see Fig.~\ref{figMCP}.
In this case, one would observe a discontinuous behavior at the
meeting point. Our numerical data do not allow us to distinguish
between the two scenarios. We only observe that, if the DO-O line
becomes eventually of first order by decreasing $K$, this should occur
very close to the meeting point.

It is interesting to observe that a first-order meeting point is 
expected when the transitions are associated with one $N$ vector
parameter $\phi_1$ and one scalar order parameter $\phi_2$, which are
both local. Indeed, the corresponding LGW
model~\cite{LF-72,FN-74,NKF-74}, with Hamiltonian
\begin{eqnarray}
L &=& {1\over 2} \sum_\mu [(\partial_\mu \phi_1)^2 + (\partial_\mu \phi_2)^2] +
    {1\over 2} (r_1 \phi_1^2 + r_2 \phi_2^2)
\nonumber \\
   && + 
     u_1 (\phi_1^2)^2 + u_2 \phi_2^4 + w \phi_1^2 \phi_2^2,
\end{eqnarray}
does not admit any fixed point for any $N\ge 2$, see, e.g.,
Refs.~\cite{CPV-03,PV-02,KAE-23,Aharony-24}.  Only the multicritical
${\rm Z}_2\oplus {\rm Z}_2$ LGW theory, corresponding to $N=1$, has a
stable bicritical fixed point belonging to the $XY$ universality
class.  This effective LGW model has been used to investigate the
nature of the transitions close to the meeting point in the ${\rm
  Z}_2$ gauge Higgs model. In that case, however, duality allowed us
to argue that the nonlocal order parameter could be mapped by duality
onto a local one. Duality is missing here and therefore the relation
between the local LGW model and our gauge model is unclear.

\section{Vector correlations in the presence of a stochastic gauge fixing}
\label{gaufix}

In this section we would like to come back to the question of the
appropriate order parameter for the DO-O transitions.  As discussed in
Sec.~\ref{Kinftytrline}, a correct LGW description requires a vector
order parameter, but this is apparently at odds with the gauge
invariance of the model. Indeed, the lattice vector field ${\bm s}_x$
is not gauge invariant and therefore its correlation functions are
trivial. In particular, its two-point function
\begin{equation}
G_s({\bm x},{\bm y})=\langle {\bm
  s}_{\bm x} \cdot {\bm s}_{\bm y}\rangle
\label{gsdef}
\end{equation}
trivially vanishes for ${\bm x}\neq {\bm y}$. This apparent puzzle can
be solved by showing that critical vector correlations can be
uncovered by an appropriate gauge fixing. For this purpose, we
implement the stochastic gauge fixing outlined in
Ref.~\cite{BPV-24-gaufix}. It leaves gauge-invariant correlations
invariant, it is thermodynamically well defined, and allows us to
unveil the critical vector modes that effectively drive the DO-O
transitions.

Because of the discrete nature of the gauge variables, standard gauge
fixing procedures cannot be applied.  Therefore the idea is to average
non-gauge invariant quantities over all possible gauge
transformations, i.e.
\begin{eqnarray}
{\bm s}_{\bm x} &\to& \hat{\bm s}_{\bm x}=w_{\bm x}{\bm s}_{\bm
    x}, \label{gautra}\\
\sigma_{{\bm x},\mu}  &\to& 
\hat{\sigma}_{{\bm x},\mu}=w_{\bm x}\sigma_{{\bm
    x},\mu}w_{{\bm x}+\hat{\mu}},
\nonumber
\end{eqnarray}
using an appropriate weight for the ${\mathbb Z}_2$ site variables
$w_{\bm x}=\pm 1$. As discussed in Ref.~\cite{BPV-24-gaufix}, a
convenient choice is provided by the Gibbs weight
$\exp[-H_w(\sigma,w)]$, with the ancillary Hamiltonian
\begin{eqnarray} 
  H_w=-\gamma \sum_{{\bm x},\mu} w_{\bm x} \sigma_{{\bm x},\mu}
  w_{{\bm x},\mu},\qquad \gamma>0,
\end{eqnarray}
so that positive values of $\hat{\sigma}_{{\bm x},\mu} = w_{\bm
  x}\sigma_{{\bm x},\mu}w_{{\bm x}+\hat{\mu}}$ are favored. 
Correspondingly, we define a gauge-fixed two-point spin function 
$\widehat{G}_s$ as
\begin{equation}
  \widehat{G}_s({\bm x},{\bm y}) = \langle [\hat{\bm s}_{\bm x} \cdot
    \hat{\bm s}_{\bm y}] \rangle = {\sum_{\{{\bm s},\sigma\}}
    e^{-H({\bm s},\sigma)} [\hat{\bm s}_{\bm x} \cdot \hat{\bm s}_{\bm
        y}] \over \sum_{\{{\bm s},\sigma\}} e^{-H({\bm s},\sigma)}}
    \label{gsaver}
\end{equation}
where $\hat{\bm s}_{\bm x}=w_{\bm x}{\bm s}_{\bm x}$, $H$ is the
gauge-invariant Hamiltonian (\ref{ham}), and
\begin{equation}
  [\hat{\bm s}_{\bm x} \cdot \hat{\bm s}_{\bm y}] = {\sum_{\{w\}}
    e^{-H_w(\sigma,w)} \hat{\bm s}_{\bm x} \cdot \hat{\bm s}_{\bm y} \over
    \sum_{\{w\}} e^{-H_w(\sigma,w)}}.
    \label{gaugeav}
\end{equation}
Here $[\cdot ]$ indicates the (quenched) average over the 
${\mathbb Z}_2$ fields with weight $e^{-H_w}$, 
for fixed values of ${\bm s}_{\bm x}$ and $\sigma_{{\bm x},\mu}$,
while $\langle \cdot \rangle$ is the standard average 
over ${\bm s}_{\bm x}$ and
$\sigma_{{\bm x},\mu}$ with the gauge-invariant weight $e^{-H}$.

Note that the resulting model with the added variables $w_{\bm x}$ is
a quenched random-bond Ising model~\cite{EA-75} ($w_{\bm x}$ are the
Ising variables), with a particular choice of bond distribution,
determined by the gauge-invariant average over the variables ${\bm
  s}_{\bm x}$ and $\sigma_{{\bm x},\mu}$ of the ${\mathbb Z}_2$-gauge
$N$-vector model.  We recall that quenched random-bond Ising models
have several phases---disordered, ferromagnetic, and glassy
phases---depending on the temperature, the amount of randomness of the
bond distribution, and its spatial correlations, see, e.g.,
Refs.~\cite{HPPV-07,HPV-07,CP-19}. In particular, we expect the
present model to undergo a quenched transition for $\gamma =
\gamma_c(J,K)$. The transition separates a disordered phase for
$\gamma < \gamma_c(J,K)$ from a large-$\gamma$ phase, which, a priori,
can be ferromagnetic or glassy, depending on the nature of the bond
coupling.

A key point of the above procedure concerns the value of $\gamma$,
which should be chosen such that the spins ${\bm s}_{\bm x}$ become
critical at the transition. For this purpose, $\gamma$ must be
large---more precisely, it should satisfy $\gamma >
\gamma_c(J,K)$---to ensure that the variables $w_{\bm x}$ are ordered,
effectively favouring positive values for the link variables
$\hat{\sigma}_{{\bm x},\mu} = w_{\bm x}\sigma_{{\bm x},\mu}w_{{\bm
    x}+\hat{\mu}}$.  On the other hand, for $\gamma < \gamma_c(J,K)$,
we do not expect vector correlations to become critical.

Quenched averages are computed as in standard simulations of random
quenched systems.  We simulate the model with Hamiltonian $H$ and every $N_s$
sweeps we compute the gauge averages over the $w_{\bm x}$ variables
for fixed values of ${\bm s}_{\bm x}$ and $\sigma_{{\bm x},\mu}$ (at
fixed disorder in the language of random systems). We use a standard
Metropolis update.  For each ``disorder realization," we perform about
$10^5$ sweeps of the whole lattice. After discarding 
approximately $O(10^4)$ sweeps to ensure thermalization, 
we perform approximately $10^2$ measurements
(this is probably
much more than needed, but it guarantees the absence of any
initialization bias).

\begin{figure}[tbp]
\includegraphics[width=0.95\columnwidth, clip]{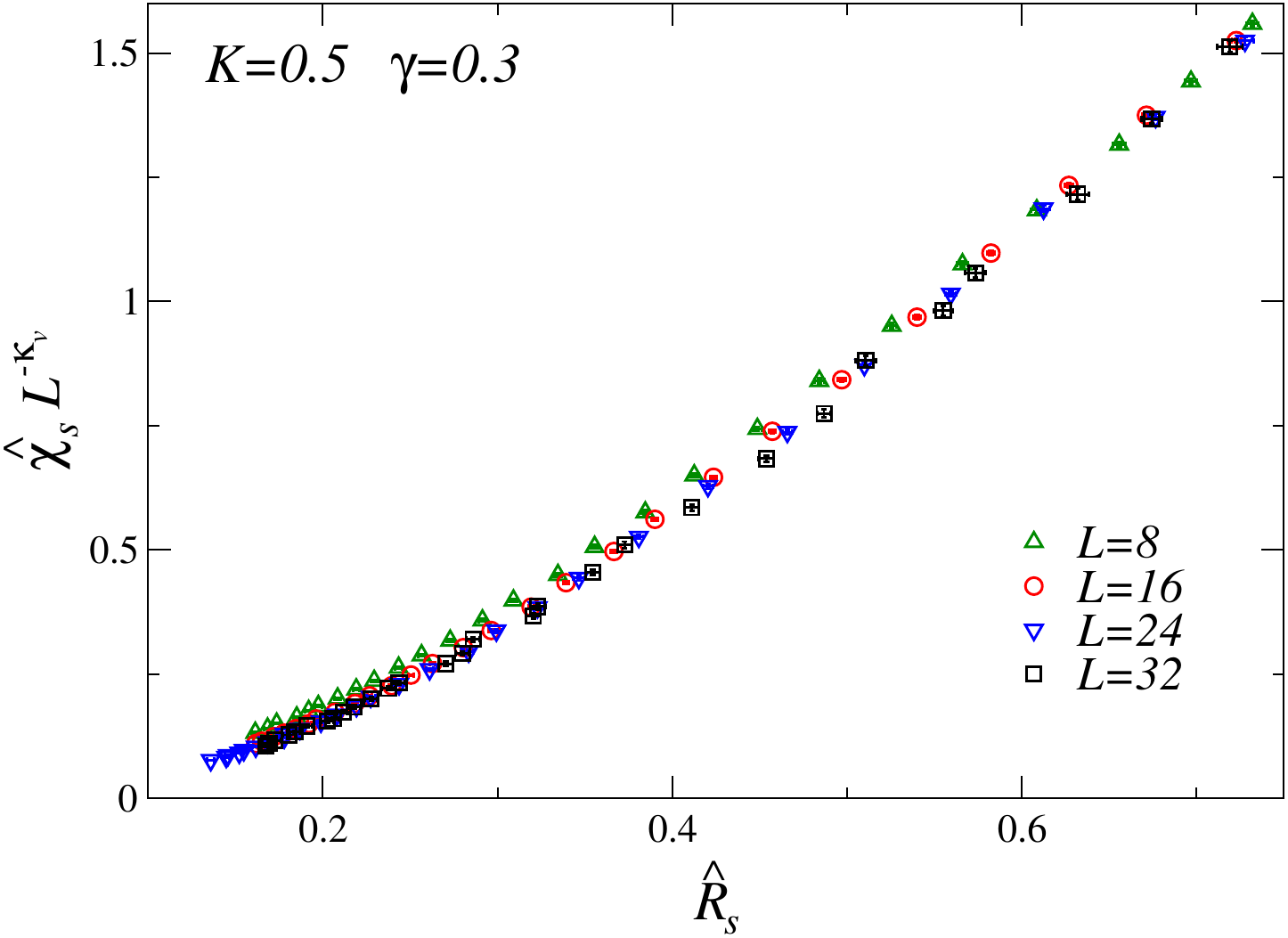}
\caption{Plot of $L^{-\kappa_v}\chi_s$ versus $\hat{R}_s=\hat\xi_s/L$
  for $K=1$, $\gamma=0.3$. Here $\hat\chi_s$ and $\hat\xi_s$ are
  defined in terms of $\widehat{G}_s$, see Eq.~(\ref{gsaver}). We set
  $\kappa_v=3 - 2Y_{V,XY} = 1.96182(2)$, where $Y_{V,XY}$ is the RG
  dimension of the vector field in the $XY$ universality class.  }
\label{figchisgf}
\end{figure}

\begin{figure}[tbp]
\includegraphics[width=0.95\columnwidth, clip]{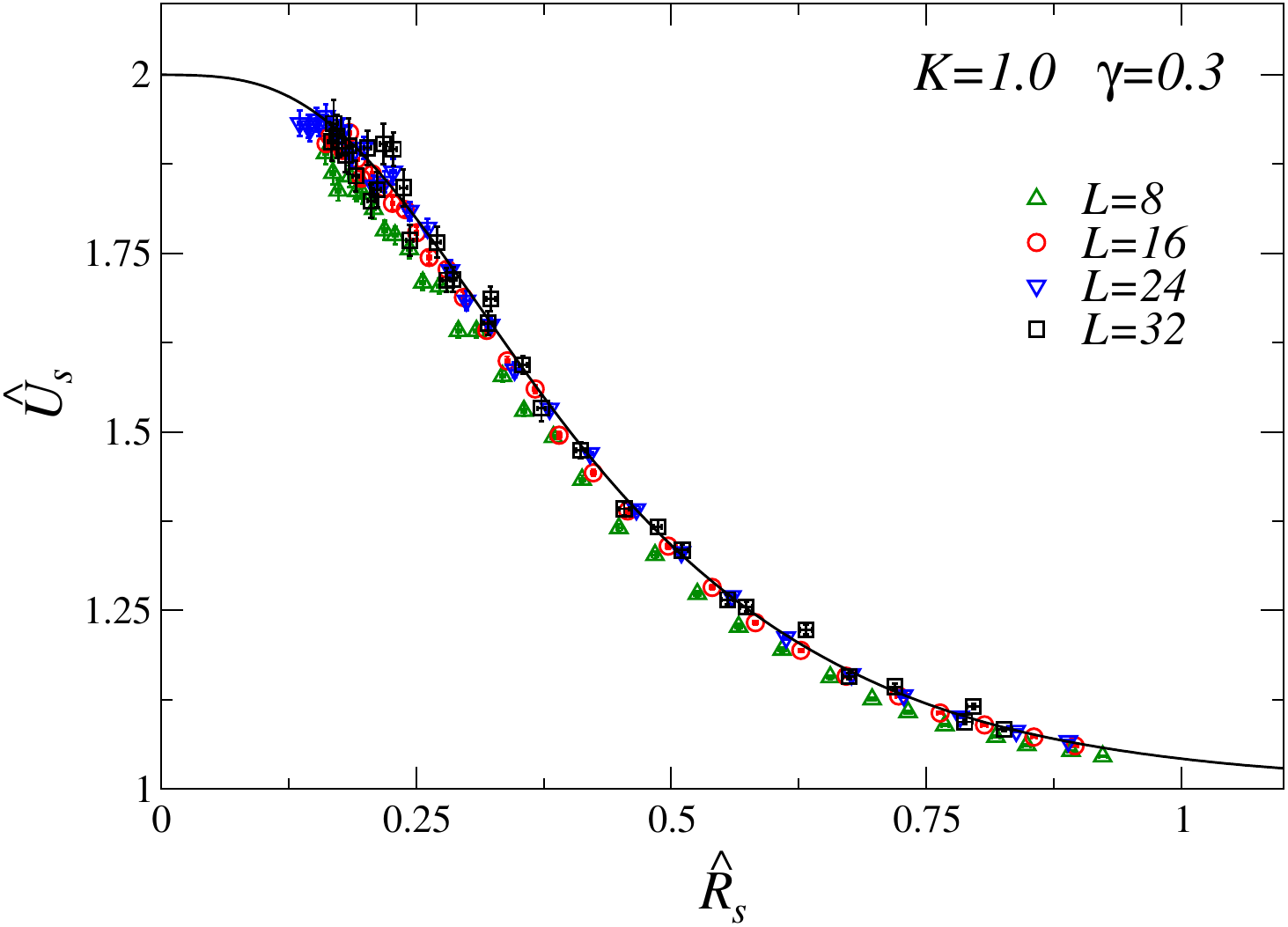}
\caption{The Binder parameter $\hat{U}_s$ as a function of
  $\hat{R}_s\equiv \hat\xi_s/L$ for $K=1$ and $\gamma=0.3$. The data
  appear to approach the universal scaling curve (solid line) for the
  $XY$ model obtained in Ref.~\cite{BPV-21-coAH}.  }
\label{figUsvsRsgf}
\end{figure}

In the following we show that, for sufficiently large values of
$\gamma$, along the DO-O transition line the two-point function
$\widehat{G}_s$ behaves as the vector correlation function in the $XY$
model.  For this purpose, it is useful to define the susceptibility
$\hat\chi_s$ and the second-moment correlation length $\hat\xi_s$,
which can be defined as in Eq.~(\ref{xidefpb}) in terms of
$\widehat{G}_s$.  We also define the Binder parameter associated with
the spin variables $\hat{\bm s}_{\bm x}$
\begin{equation}
  \hat{U}_s = \frac{\langle [m_{2s}^2]\rangle}{\langle [m_{2s}]\rangle^2},
  \qquad m_{2s} = \frac{1}{L^d} \sum_{{\bm x},{\bm y}} \hat{\bm
    s}_{\bm x}\cdot \hat{\bm s}_{\bm y}.
\end{equation}
We show now results at the transition for $K=1$. We first verified that the
model with $K=1$ and $J=J_c$ has a transition for $\gamma = \gamma_c \approx
0.22$, which separates a disordered small-$\gamma$ phase from a
ferromagnetically ordered large-$\gamma$ phase.  We thus fixed $\gamma = 0.3$.
To verify that such value corresponds to a ferromagnetic ordered phase, we have
considered the Binder parameter for the overlap of the variables $w_{\bm x}$,
which appears to approach the value $U_w=1$ as $L\to \infty$, with
inverse-volume corrections, as expected for a ferromagnetic phase.

In Fig.~\ref{figchisgf} we show the results for the susceptibility
$\hat\chi_s$ for $K=1$ and $\gamma=0.3$. They demonstrate that it behaves
as the $XY$ vector susceptibiliy. Indeed, we find that $\hat\chi_s\sim
L^{\kappa_v}$ with $\kappa_c=3 - 2 Y_{V,XY}=2-\eta_{XY}$.  This is
also confirmed by the plot of $\hat{U}_s$ versus $R_s\equiv
\xi_s/L$ reported in Fig.~\ref{figUsvsRsgf}. Indeed, the data
converge toward the corresponding $XY$ universal curve. These results
are expected to hold for any value of $\gamma$, as long as
$\gamma>\gamma_c\approx 0.22$ for $K=1$.

Along the small-$K$ DD-O line the data suggest a discontinuous
behavior of $\widehat{G}_s$ for $\gamma\gtrsim 0.3$. 
This result is consistent with the general picture. Indeed, the results
along the DO-O line indicate that vector modes magnetize as 
$J$ increases across the DO-O line. Physically, we do not expect 
the gauge-fixing procedure to give rise to additional transition lines
in the O phase. Therefore, vector modes should also magnetize 
as $J$ increases across the DD-O line. However,
along the DD-O line, the operator $Q_{\bm x}^{ab}$ is the order 
parameter which behaves as a vector $XY$
field, so there cannot be an additional emerging critical vector
field. Thus, $\widehat{G}_s$ is discontinuous, but not critical.

We finally remark that analogous results are expected at the O($N$)
continuous transitions along the DO-O transition line for higher
values of $N$. See in particular Ref.~\cite{BPV-24-gaufix} for other
applications of the stochastic gauge fixing.

\section{Conclusions}
\label{conclu}

We have investigated the phase diagram and critical behavior of 3D
lattice ${\mathbb Z}_2$-gauge $N$-vector models. Their Hamiltonian
Eq.~(\ref{ham}) is obtained by minimally coupling $N$-component real
variables with ${\mathbb Z}_2$ gauge variables, with a global O($N$)
and local ${\mathbb Z}_2$ invariance.  They represent
paradigmatic models with different phases characterized by the
spontaneous breaking of the global O($N$) symmetry and by the
different topological properties of the ${\mathbb Z}_2$-gauge
excitations.

The 3D ${\mathbb Z}_2$-gauge $N$-vector model presents three phases
for any $N\ge 2$, distinguished by the order/disorder of the spin
correlations and the order/disorder of the ${\mathbb Z}_2$-gauge
correlations, see Fig.~\ref{phadiaN}. These phases are separated by
three transition lines.

(i)~At small $J$, the small-$K$ and large-$K$
spin-disordered phases are separated by a line of topological
transitions. Their continuous transitions belong to the ${\mathbb
  Z}_2$-gauge universality class for any $N$.

(ii)~The transitions along the small-$K$ DD-O line are first order for
any $N\ge 3$. For $N=2$ we observe continuous transitions for
sufficiently small $K$, belonging to the 3D $XY$ universality class;
they turn into first-order ones as $K$ is increased, before reaching
the meeting point.

(iii)~The transitions along the large-$K$ DO-O line are expected to be
continuous for any $N$ (at least for sufficiently large $K$). These
transitions belong to the O($N$) vector universality class.  It is
important to note that these O($N$) transitions are quite peculiar,
since critical vector correlations emerge only after an
appropriate gauge fixing and, in particular, when the stochastic gauge
fixing outlined in the companion paper \cite{BPV-24-gaufix} is used.  This
gauge fixing 
approach is thermodynamically consistent and local, thus it allows us
to apply standard RG arguments to the stochastically gauge-fixed theory.

In this work we mainly focus on models with $N=2$. At variance with
what happens when $N\ge 3$, in this case the small-$K$ DD-O
transitions can be continuous---for larger values of $N$ they are of
first order. Interestingly, the small-$K$ DD-O transitions and the
large-$K$ DO-O transitions both belong to the $XY$ universality
class. In spite of this, the critical behavior along the two lines is
different. Indeed, for small values of $K$ the vector $XY$ order
parameter is gauge invariant.  Instead, for large values of $K$, the
model has an emergent order parameter, the spin ${\bm s}_{\bm x}$
within an appropriate stochastic gauge fixing because it is not
gauge-invariant.  The different nature of the critical modes can be
probed by studying the correlations of the gauge-invariant operator
$Q_{\bm x}^{ab}$. Along the small-$K$ DD-O transition line, its RG
dimension $Y_Q$ coincides with the RG dimension
$Y_{V,XY}=0.519088(22)$ of the vector field in the $XY$ universality
class.  On the other hand, along the large-$K$ DO-O line, since the
order parameter is the spin $s_{\bm x}$, $Q_{\bm x}^{ab}$ behaves as a
tensor spin-2 operator. Therefore, we predict $Y_Q = Y_{T,XY}$ where
$Y_{T,XY}=1.23629(11)$, where $Y_{T,XY}$ is the spin-two RG dimension
in the $XY$ universality class.  This different behavior is easily
detected by studying the size dependence of the corresponding
susceptibility, which diverges as $\chi\sim L^{1.9618}$ along the
small-$K$ DD-O transition line and as $\chi\sim L^{0.5274}$ along the
large-$K$ DO-O transition line.

To verify the previous predictions we have performed MC simulations
for $N=2$, see Table~\ref{tabres}. The FSS analyses of the data
confirm the general results presented above and, in particular, the
two different effective $XY$ descriptions of the small-$K$ DD-O and
large-$K$ DO-O transition lines.

\acknowledgments

The authors acknowledge support from project PRIN 2022 ``Emerging
gauge theories: critical properties and quantum dynamics''
(20227JZKWP).  Numerical simulations have been performed on the CSN4
cluster of the Scientific Computing Center at INFN-PISA.

\end{document}